\documentclass{aa}
\usepackage{graphicx,natbib}
\usepackage[usenames]{color}

\bibpunct{(}{)}{;}{a}{}{,}

\def\apj{ApJ}

\def\flux{\rm erg~ s$^{-1}$~ cm$^{-2}$}
\def\lum{\rm erg~ s$^{-1}$}

\begin{document}

\sloppypar

   \title{Universal X-ray emissivity of the stellar population in
early-type galaxies: unresolved X-ray sources in NGC~3379 
}

   \author{M.~Revnivtsev \inst{1,2},  E.~Churazov \inst{1,2}, S.~Sazonov \inst{1,2}, W.~Forman \inst{3}, C.~Jones \inst{3}
}

   \offprints{mikej@mpa-garching.mpg.de}

   \institute{
              Max-Planck-Institute f\"ur Astrophysik,
              Karl-Schwarzschild-Str. 1, D-85740 Garching bei M\"unchen,
              Germany,
     \and
              Space Research Institute, Russian Academy of Sciences,
              Profsoyuznaya 84/32, 117997 Moscow, Russia
       \and
               Harvard-Smithsonian Center for Astrophysics,
               60 Garden Street, Cambridge, MA 02138, USA
            }
  \date{}

        \authorrunning{Revnivtsev et al.}
        \titlerunning{Unresolved X-ray sources in NGC~3379 }

\abstract{We use deep {\em Chandra} observations to measure the
  emissivity of the unresolved X-ray emission in the elliptical galaxy
  NGC~3379. After elimination of bright, low-mass X-ray binaries with
  luminosities $\ga 10^{36}$ \lum, we find that the remaining
  unresolved X-ray emission is characterized by an emissivity per unit
  stellar mass $L_{\rm x}/M_\ast\sim8.2\times10^{27}$
  \lum\ M$^{-1}_{\odot}$ in the 0.5--2~keV energy band. This value is
  in good agreement with those previosuly determined for the dwarf
  elliptical galaxy M32, the bulge of the spiral galaxy M31 and the
  Milky Way, as well as with the integrated X-ray emissivity of
  cataclysmic variables and coronally active binaries in the Solar
  neighborhood. This strongly suggests that i) the bulk of the
  unresolved X-ray emission in NGC~3379 is produced by its old stellar
  population and ii) the old stellar populations in all galaxies can
  be characterized by a universal value of X-ray emissivity per unit
  stellar mass or per unit $K$ band luminosity. 
  }

   \maketitle

%

\section{Introduction}

Non-active galaxies  produce X-ray emission
that generally consists of a diffuse component originating in hot
interstellar gas and emission from point-like sources
(e.g. \citealt{tf85}). The diffuse X-ray emission is a powerful tool
that provides a record of past star formation and central massive
black hole activity through their effects on the interstellar medium
(e.g. \citealt{heckman90,lm91,strickland00,churazov01,birzan04}), and to measure total
(including dark matter) masses of elliptical galaxies
\citep{forman85}.

Luminous ($L_{\rm x}\ga 10^{35}$--$10^{36}$~erg~s$^{-1}$) compact
sources, including low- and high-mass X-ray binaries (LMXBs and HMXBs,
respectively) can be individually detected in nearby galaxies with
{\sl Chandra} and {\sl XMM-Newton}. Since the measured luminosity
functions of both LMXBs and HMXBs are significantly flatter than
$dN/dL_{\rm x}\propto L_{\rm x}^{-2}$ at low luminosities
\citep{gilfanov04}, many studies assumed that any remaining unresolved
X-ray emission in galaxies must originate in hot interstellar gas
\cite[e.g.][to name but a few]{shirey01,strickland04}.

However, it was recently recognized that weak X-ray sources with
luminosities $10^{28}$--$10^{34}$~erg~s$^{-1}$ located in the Solar
vicinity, in particular cataclysmic variables (CVs) and coronally
active binaries (ABs), together produce a considerable X-ray
luminosity per unit stellar mass \citep{sazonov06}, which allows one
to explain the apparently diffuse X-ray emission distributed over our
Galaxy (the so-called Galactic Ridge X-ray Emission) as the
superposition of a great number of such low-luminosity sources
\citep{mikej06}. Since CVs and ABs (and LMXBs) are not associated
with ongoing or recent star formation, it is reasonable to expect that
such objects should produce, in {\em all} galaxies, approximately
the same amount of X-ray emission per unit stellar mass as in the
Milky Way.

A particularly important contribution from unresolved point sources
may be expected in the soft X-ray band (0.5--2~keV). Indeed, the
cumulative emissivity of low-luminosity sources in the Solar
neighborhood is estimated at $dL_{\rm x}/M_\ast\sim$1--3$\times
10^{28}$~erg~s$^{-1}$ \citep{sazonov06}, which
for a galaxy of mass $M_\ast$ implies a total luminosity of a few $\times 
10^{39}(M_\ast/10^{11}M_\odot)$~\lum\ in unresolved soft X-ray
emission, which is just a few times smaller than the expected cumulative
luminosity of bright LMXBs \citep{gilfanov04}. In fact, the possibility that 
unresolved point sources might provide an important contribution to the 
X-ray luminosity of galaxies was discussed in several early works
\citep{pellegrini94,irwin98} but was not supported by quantitative
estimates \cite[see however some estimates in][]{pellegrini07}.
 Importantly, both in coronally active stars and in
cataclysmic variables (i.e. accreting white dwarfs), we 
are dealing with thermal X-ray emission from an optically thin plasma,
with temperatures $10^{6}$--$10^{7}$~K and $10^{7}$--$10^{8}$~K,
respectively, similar to that produced by a hot, multi-temperature
interstellar gas. In particular, X-ray spectra of coronal stars 
show a prominent peak at energies 0.6--0.8~keV due to a blend of
atomic lines, and the same is expected for interstellar gas with
temperatures $\sim$0.5--1~keV. Therefore, 
separating the collective X-ray emission of faint point sources from
truly diffuse emission produced by hot interstellar gas is hardly
possible by means of X-ray spectroscopy alone. 

One might hope, however, to be able to predict the contribution of
unresolved point sources to the X-ray emission of a given galaxy using
the measured cumulative luminosity (per unit stellar mass) of faint
X-ray sources in the Solar vicinity. But how well does this locally
determined value characterize other galaxies? Is the cumulative X-ray
emissivity of faint sources strongly sensitive to the properties (age,
metallicity, etc.) of the stellar population of a given galaxy? To
answer these questions, we selected several nearby galaxies with
available {\sl Chandra} observations that are likely to be
sufficiently gas poor for the bulk of their unresolved X-ray emission
to be due to unresolved point sources. The first galaxy studied was
the dwarf elliptical galaxy M32, for which we demonstrated
\citep{mikej_m32} that the soft X-ray emissivity (with bright LMXBs
excluded) per unit stellar mass is compatible with the combined
emissivity of CVs and ABs in the Solar vicinity \citep{sazonov06}.

\begin{figure*}
\centerline{
\includegraphics[width=0.34\textwidth]{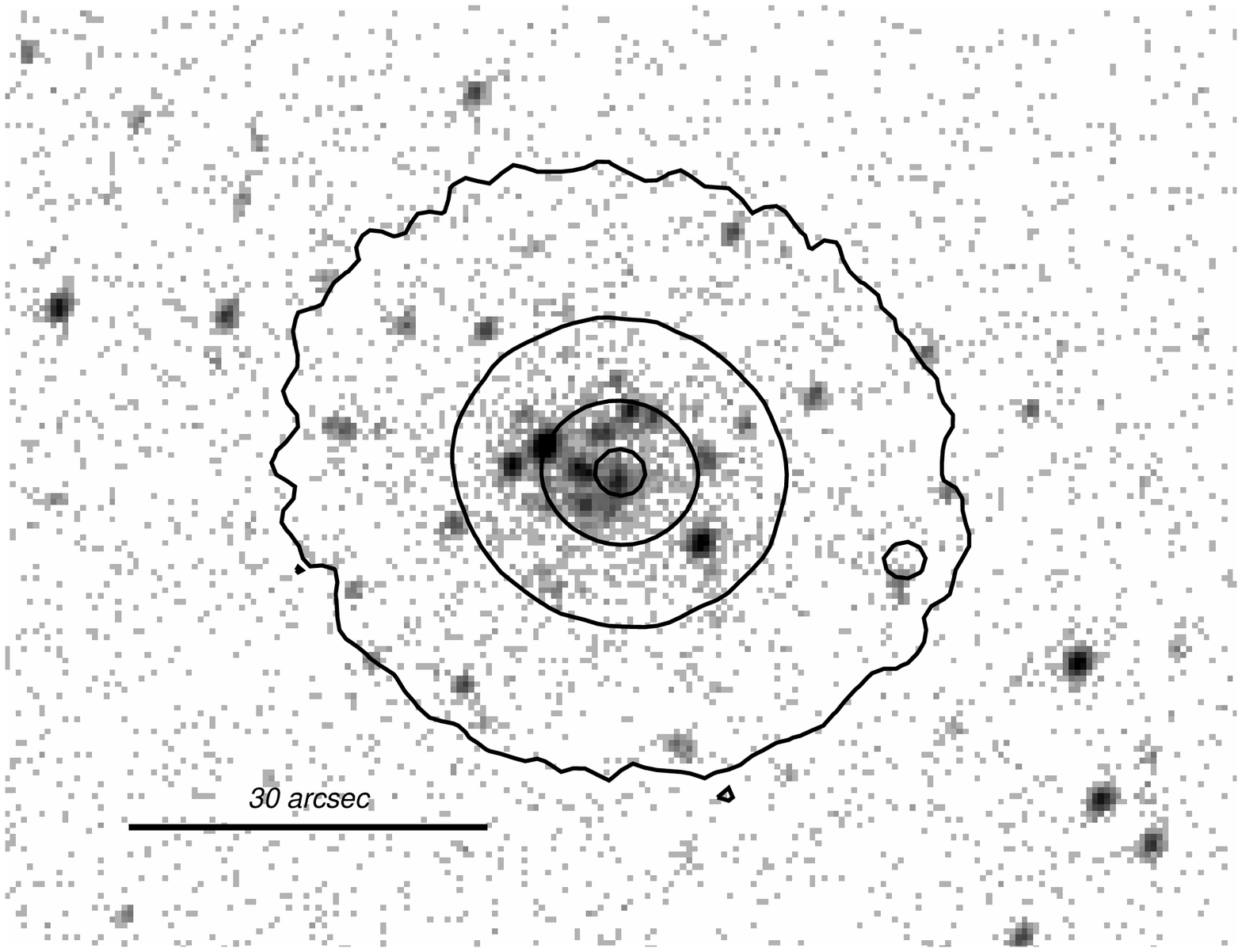}
\includegraphics[width=0.315\textwidth,bb=308 286 574 507,clip]{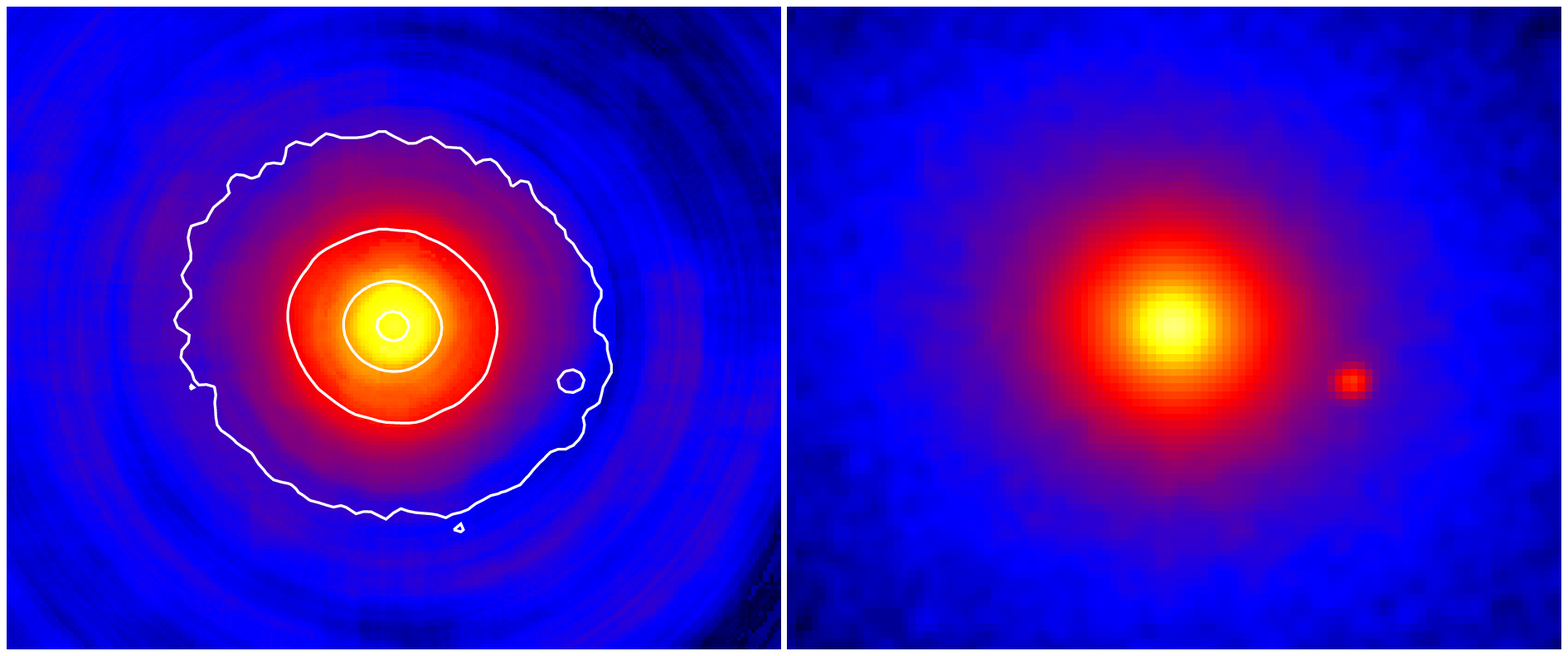}
}
\caption{Raw {\em Chandra} image (left panel) of NGC~3379 compared with the
near-infrared ($K$-band) image of the galaxy (right panel). Contours
in the left image are logarithmically spaced
near-infrared brightness isophotes.  
}
\label{images}
\end{figure*}

In the present paper, we extend this study to a moderately massive
elliptical galaxy, NGC~3379. We focus on studying the properties of
the unresolved X-ray component in the soft band 0.5--2~keV, where the
sensitivity of {\em Chandra} is maximal. We also discuss the
broad-band spectral properties of this emission and the implications
for early-type galaxies in general. The unresolved X-ray component of
NGC~3379 was previously studied by \cite{david05} and \cite{fukazawa06} 
using smaller datasets.  Application of the usual assumption that the
hot X-ray emitting interstellar gas is in hydrostatic equilibrium in this galaxy implies
an unusually high mass-to-light ratio, which led to the suggestion
\cite[e.g.][]{fukazawa06} that a significant contribution to the observed
X-ray flux of the galaxy comes from unresolved point sources. 
We argue in this paper that the bulk of this emission is
likely the superposition of faint point sources, CVs and ABs.

We adopt a distance to NGC~3379 of 9.8~Mpc \citep{jensen03}.

\section{Data analysis}

We used all publicly available {\em Chandra}
observations of NGC~3379 (ObsIDs 1587, 4692, 7073, 7074, 7075, 7076),
which in total provide approximately 340~ks of useful exposure time.

The \emph{Chandra} data were reduced following a standard procedure
fully described in \cite{2005ApJ...628..655V}. The detector
background was modeled using the stowed dataset 
(http://cxc.harvard.edu/contrib/maxim/stowed). To additionally check
the accuracy of the background subtraction, we also measured 
the detector background in regions away from NGC~3379. 

Point sources were detected in the 0.5--7~keV energy band using the
wavelet decomposition package $wvdecomp$ of $ZHTOOLS$
\citep{vikhlinin98}\footnote{http://hea-www.harvard.edu/saord/zhtools/}.
To study the unresolved X-ray emission of NGC~3379, we masked regions
with radii $2.46\arcsec$ around the detected point sources, which
ensures removal of more than 98\% of the flux from these sources in
the region of our study ($<100$\arcsec\ around the aimpoint of
\emph{Chandra}).

In our the spectral analysis, we adopted a Galactic photoabsorption
column density toward NGC~3379 $n_H L=2.7\times10^{20}$
cm$^{-2}$ \citep{dickey90} and all fluxes quoted below are 
corrected for this absorption.

\section{Unresolved X-ray emission in NGC~3379}

A conservative source detection threshold of 10 net counts corresponds for
the analyzed (relatively long exposure) dataset to a flux
$\sim 10^{-16}$ \flux\ in the energy band 0.5--7~keV. Given the
adopted NGC~3379 distance, this allows us to detect and remove point
sources with luminosities higher than $\sim2\times10^{36}$ \lum. 

\subsection{Properties of the unresolved X-ray emission} 

In Fig.~\ref{images} we compare the raw {\em  Chandra} image of NGC~3379,
 and the near-infrared
($K$-band) image of the galaxy obtained from the 2MASS Large Galaxy
Atlas \cite[][see also
  http://irsa.ipac.caltech.edu/applications/2MASS/LGA/]{jarrett03}.
In the following analysis, we assumed that the total brightness of the 
galaxy is $K=6.27$ 
(taken from the 2MASS Large Galaxy Atlas). For a distance modulus
$m-M=29.96$ \citep{jensen03}, the total near-infrared
luminosity is $L_K=6.8\times10^{10} L_{\odot}$. Using the color dependent
mass-to-light ratio $M/L_K=0.84$ determined for $B-V=0.96$ from
\cite{bell03}, we further estimated the stellar mass of NGC~3379 at
$M_\ast\approx5.7\times10^{10} M_{\odot}$. 

\begin{figure}
\includegraphics[width=\columnwidth]{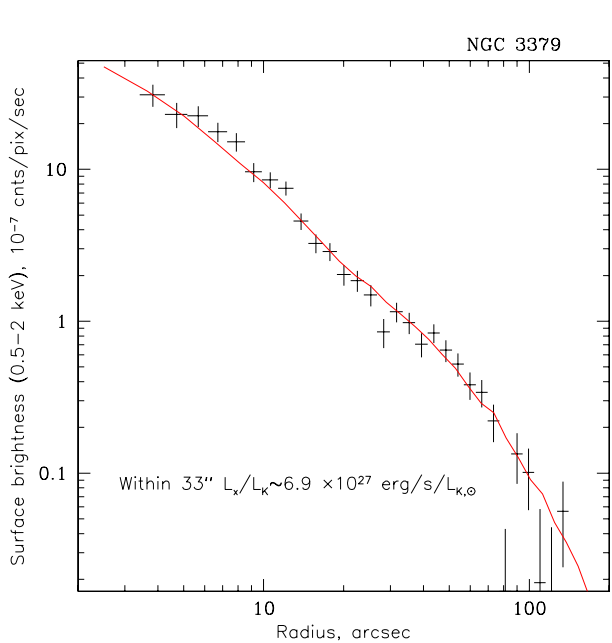}
\caption{Radial surface brightness profile of NGC~3379 in the
0.5--2~keV energy band. The contribution of point sources with luminosities
$>2\times10^{36}$ \lum\  has been subtracted. The
solid line shows the radial profile of near-infrared ($K$-band)
surface brightness, rescaled by the factor $L_{\rm
  0.5-2~keV}/L_K=6.9\times10^{27}$ \lum\ $L^{-1}_{K,\odot}$. The
agreement between the X-ray and $K$-band profiles is excellent and
supports the stellar origin of the soft X-ray emission.
}
\label{profile}
\end{figure}

\begin{figure}
\includegraphics[width=\columnwidth,bb=37 180 570 530,clip ]{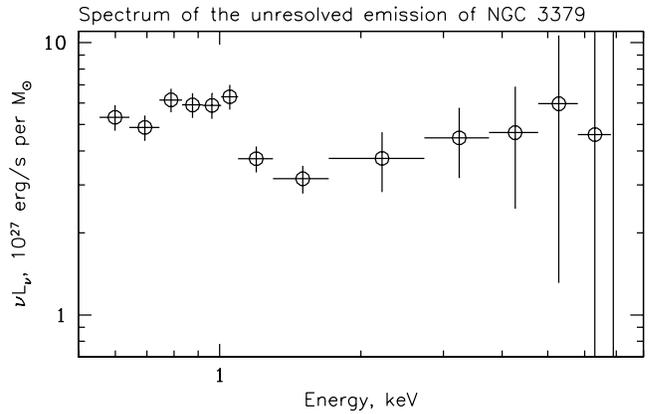}
\caption{Spectrum of the unresolved X-ray emission of NGC~3379, 
normalized to the stellar mass contained in the 33\arcsec
central region. 
}

\label{spectrum}
\end{figure}

In Fig.~\ref{profile} we compare the radial surface brightness
profiles of the unresolved X-ray emission (i.e., excuding detected
point sources) with the near-infrared light of NGC~3379. In
Fig.~\ref{spectrum}, we show the energy spectrum of the unresolved
X-ray emission collected by {\em Chandra} within the central region of
NGC~3379 of radius 33\arcsec, which is approximately the effective
radius of the galaxy in the near-infrared spectral band.

One can see from Fig.~\ref{profile} that 
the unresolved X-ray (0.5--2~keV) emission very closely follows 
the near-infrared light in NGC~3379, which implies that this emission is
characterized by a constant ratio of X-ray and $K$-band emissivities,
or by constant X-ray emissivity per unit stellar mass. In the region
$<33 \arcsec$ around the center of the galaxy:
\begin{equation}
{L_{\rm 0.5-2~keV}\over{L_K}}=(6.9\pm0.7)\times10^{27}~{\rm
erg^{-1} s^{-1}} L^{-1}_{K,\odot},
\label{lx_lk}
\end{equation}
\begin{equation}
{L_{\rm 0.5-2~keV}\over{M_\ast}}=8.2\pm0.8(\pm2.5)\times10^{27}
~{\rm erg^{-1} s^{-1}} M^{-1}_{\rm \odot}.
\label{lx_m}
\end{equation}

The quoted uncertanties in equations~(\ref{lx_lk}) and (\ref{lx_m})
are derived from the statistical errors in the measured X-ray
flux. In addition, we assumed a 30\% uncertainty in the $L_{\rm K}$ to
$M_\ast$ conversion (see e.g. \citealt{bell03}); the corresponding
systematic uncertainty is given in parentheses in equation~(\ref{lx_m}).

\subsection{Remaining contribution of LMXBs below the detection threshold?}

\begin{figure}[htb]
\includegraphics[width=\columnwidth]{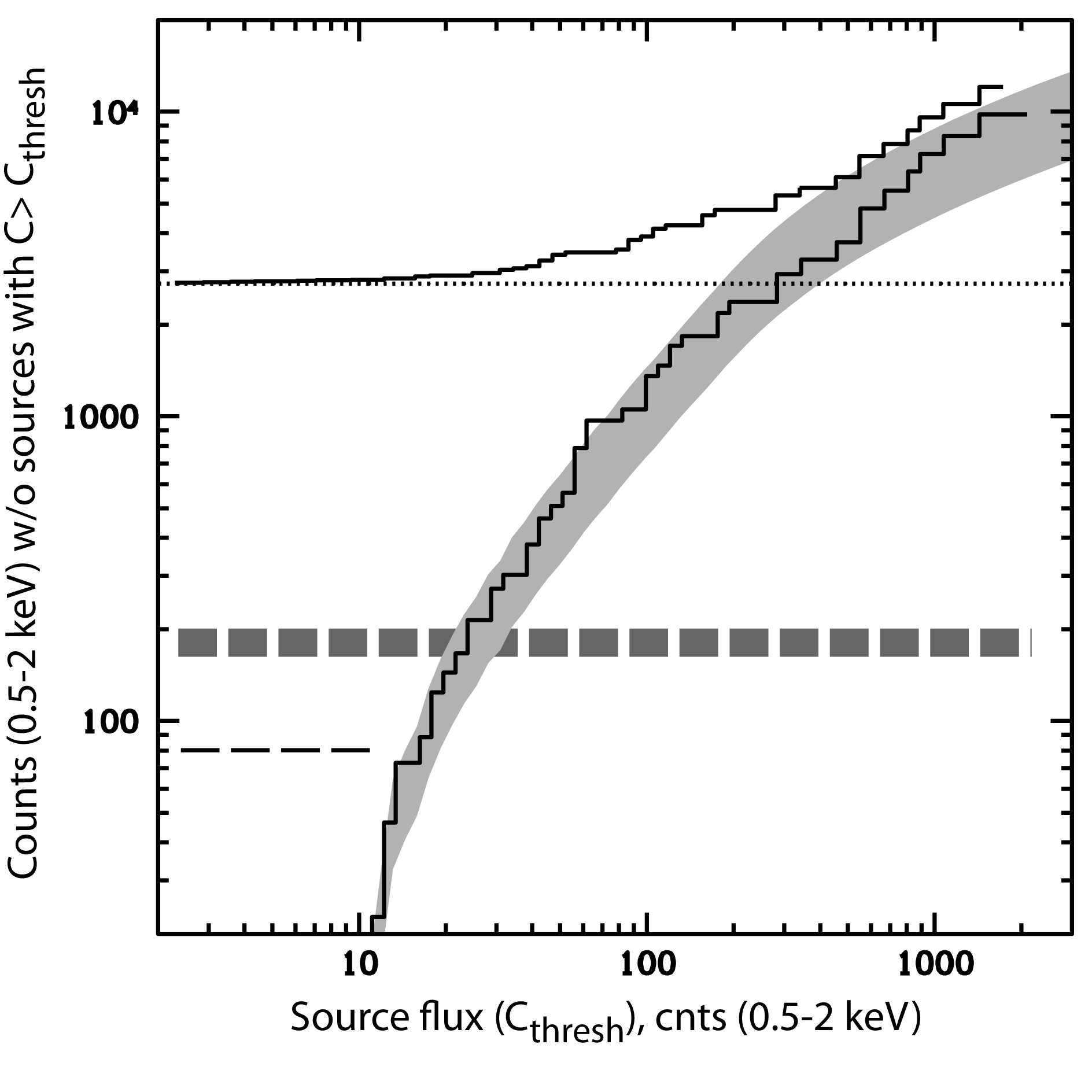}
\caption{Number of counts detected by {\em Chandra} in
  the 0.5--2~keV energy band within the central 33\arcsec~ radius
  region of NGC~3379 as a function of the threshold above which
  point sources were masked (upper solid histogram). The lower
  histogram shows $N_{thresh}$, the cumulative number of counts from
  detected point sources (i.e. those brighter than 10 counts) that are
  weaker than the given threshold. The gray area shows the predicted
  value of $N_{thresh}$ assuming that all the detected sources are
  LMXBs and is
  based on the luminosity function of LMXBs in nearby galaxies
  integrated above $L_{\rm x}=2\times10^{36}$~\lum; the thickness of
  the gray band corresponds to the variance in the normalization of
  the LMXB luminosity function between different galaxies
  \citep{gilfanov04}. Note
  that an LMXB with a typical power-law spectrum with photon index
  $\Gamma=1.6$ \citep{irwin03} providing 10 net \emph{Chandra} counts
  in the energy band 0.5--2~keV will have an X-ray luminosity $\sim
  2\times10^{36}$ \lum\ in the energy band 0.5--7~keV. The thin dashed
  line shows the expected flux of LMXBs with luminosities
  $10^{35}$--$2\times10^{36}$~\lum.  The thick gray horizontal dashed
  line denotes the residual flux outside the 2.46\arcsec radius
  circles used to mask bright sources.}
\label{resflux}
\end{figure}

Since NGC~3379 is an elliptical galaxy with virtually no ongoing star
formation, essentially all of the detected bright X-ray sources (with
luminosities above $\sim2\times10^{36}$ \lum) are expected
to be low mass X-ray binaries. Could the remaining unresolved
emission be due to LMXBs with luminosities below our detection
threshold? To investigate this possibility, we allowed the
threshold for masking point sources to vary and calculated the
total number of counts in the masked 0.5--2~keV {\em Chandra}
image as a function of the threshold (Fig.~\ref{resflux}). One can see
that this dependence flattens  at luminosities below 
$\sim2\times 10^{37}$ \lum, confirming the presence of unresolved
X-ray emission. 

If we assume that the average luminosity function of
LMXBs in nearby galaxies \citep{gilfanov04} continues below $10^{36}$
\lum\ with the same slope as measured in the range $10^{36}$--$10^{37}$ 
\lum, we find (see Fig.~\ref{resflux} and also Fig.~6 in
\citealt{mikej_m32}) that unresolved LMXBs ($L_{\rm 0.5-7~keV}\la
2\times 10^{36}$ \lum), whose collective emissivity per unit stellar mass is
estimated to be $\la 10^{27}$ \lum\ $M^{-1}_{\odot}$, cannot account
for more than $\sim$3\% of the unresolved X-ray emission of
NGC~3379.  We also estimate that less than $\sim$8\% of the
unresolved emission can result from our incomplete removal of bright 
sources (see Fig.~\ref{resflux}).

\subsection{Nature of the unresolved emission -- superposition of weak
sources}

\begin{figure}
\includegraphics[width=\columnwidth]{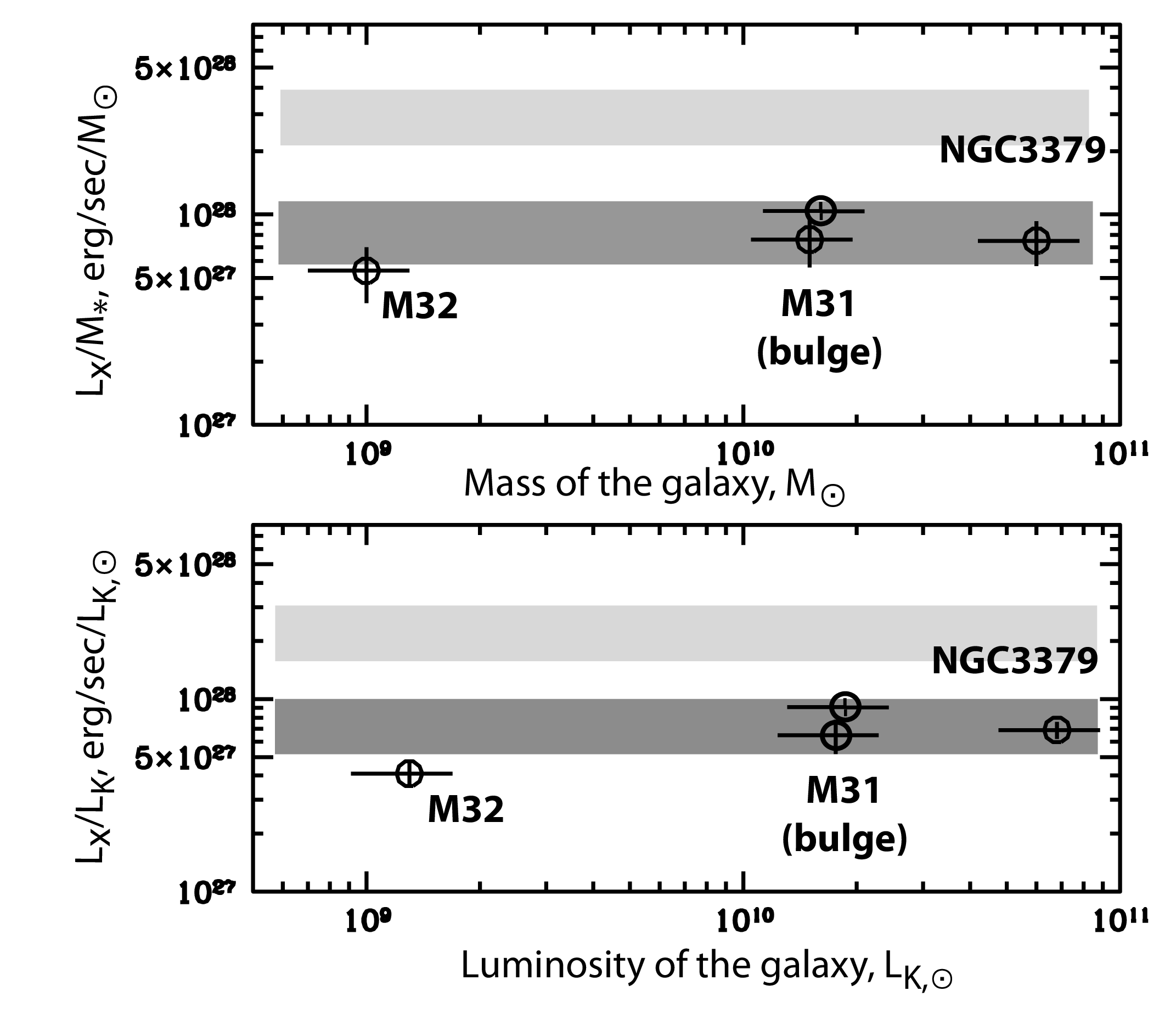}
\caption{Emissivities per unit stellar mass (upper panel) and per unit
  $K$-band luminosity (lower panel) of the unresolved soft X-ray
  (0.5--2~keV) emission in three galaxies: M32, the bulge of M31 and
  NGC~3379 (see text). 
  }
\label{ratios}
\end{figure}

What is then the origin of the unresolved soft X-ray (0.5--2~keV) 
emission of NGC~3379? Previous studies assumed that it is produced by
hot interstellar gas \citep[e.g.][]{david05,fukazawa06,pellegrini06}
and did not consider the possibility that weak ($L_{\rm x}\la 10^{35}$
\lum) undetectable X-ray sources might provide a significiant or
dominant contribution.

We recently demonstrated \citep{mikej_m32} that the unresolved X-ray
emission of the nearby dwarf elliptical galaxy M32 has the same
morphology as its near-infrared light and the X-ray emissivity per
unit stellar mass proved to be comparable to the cumulative
emissivity of faint X-ray sources -- cataclysmic variables and
coronally active binaries -- in the solar neighborhood
\citep{sazonov06}. Similarly, the unresolved X-ray emission in the
ridge of our Galaxy \citep{mikej06} and in the bulge of M31 \citep{li07} have comparable X-ray
emissivities. The similarity of these
unresolved X-ray emissivities (per solar mass) strongly suggests that
their unresolved X-ray emission arises from numerous faint X-ray
sources,  mostly CVs and ABs, and that this emission is characterized
by a nearly constant emissivity per unit stellar mass, similarly to the
case of brighter X-ray sources -- LMXBs.

The qualitatively and quantitatively similar results presented here
for NGC~3379 lead us to conclude that also in this galaxy the bulk of
the unresolved X-ray emission is produced by weak X-ray sources, mostly
CVs and ABs, rather than by hot interstellar gas. 

Figure~\ref{ratios} demonstrates that the emissivities of the
unresolved soft X-ray emission in the elliptical galaxies NGC~3379 and
M32 and in the bulge of the spiral galaxy M31 agree within a factor of
2 with each other and are compatible with the cumulative emissivity of
CVs and ABs in the Solar neighborhood. For this figure the values 
for M32 and NGC~3379 are adopted from
  \cite{mikej_m32} and the current paper, respectively. The $L_{\rm
    x}/M_{\ast}$ values for the M31 bulge are adopted from \cite{li07}
and \cite{bogdan08}
  and their conversion to $L_{\rm x}/L_K$ assumed the mass-to-light
  ratio $M_{\ast}/L_K=0.85$. The dark-gray stripe indicates the value
  and uncertainty of the cumulative X-ray emissivity of cataclysmic
  variables and coronally active binary stars (both representing the
  old stellar population) in the Solar vicinity \citep{sazonov06}; the
  estimate shown by the light-gray area also includes the contribution
  of nearby young ($\ll 1$~Gyr) coronal stars \citep{sazonov06}, which
  are expected to be rare in early-type galaxies and galactic
  bulges. The conversion from $L_{\rm x}/M_{\ast}$ to $L_{\rm x}/L_K$
  for the Solar vicinity values was done assuming $M_{\ast}/L_K=0.85$.

The spectral shape of the unresolved X-ray emission of NGC~3379
is broadly compatible with the combination of spectra of CVs 
and ABs in proportion, adopted from the Solar vicinity analysis by 
\cite{sazonov06}.However, our knowledge of the relative contributions 
of CVs and
ABs to the galactic emission in the 0.5-2.0 keV energy band 
remains limited with considerable uncertainties
 \cite[see e.g.][]{sazonov06}.

We note here that in the
Solar vicinity an additional significant soft X-ray emission is
produced by young coronal stars associated with recent star formation
and the local estimate, including this contribution, is a factor of
2--3 larger than the emissivities found for M32, NGC~3379 and for the
M31 bulge (although there is a large uncertainty in the local value,
\citealt{sazonov06}). The lower emissivity of the three early-type
galaxies/bulges is expected due to their predominantly old stellar
populations, with a much lower abundance of young stars than in the
Solar neighborhood.

\section{Universality of the X-ray emissivity of the old stellar population}

\begin{figure*}[htb]
\includegraphics[height=\textwidth,bb=167 16 444 793,clip,angle=-90]{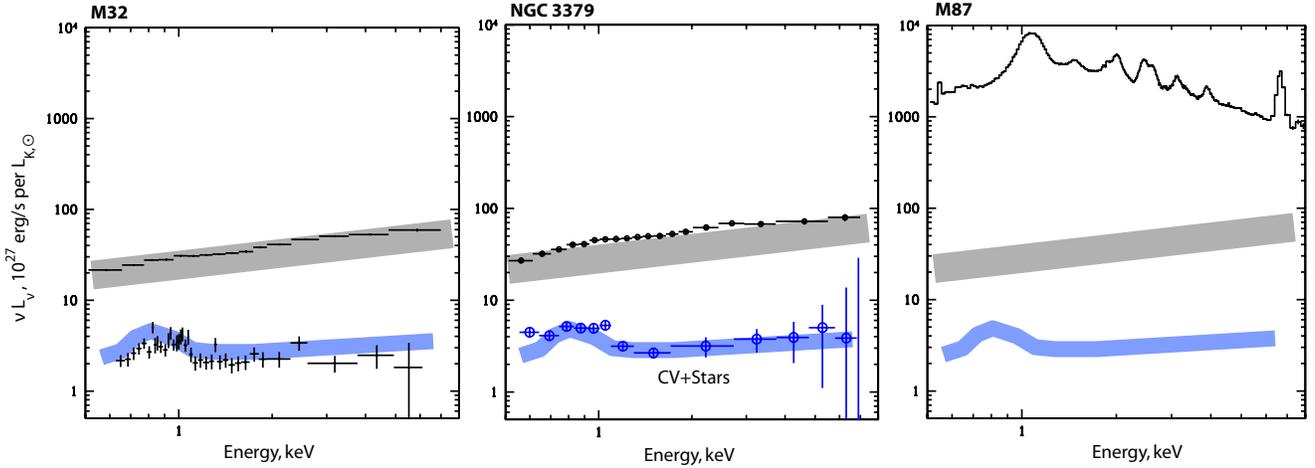}
\caption{Spectra of three elliptical galaxies of different masses
  normalized to their near-infrared ($K$-band) luminosities. The upper
  sets of {\em Chandra} data points represent the total X-ray
  emission, collected within the effective radii of the galaxies
  ($\sim 35$\arcsec\ for M32, $\sim 33$\arcsec\ for NGC~3379 and
  $\sim100$\arcsec\ for M87). The lower set of data points for M32 and
  NGC~3379 represent their unresolved X-ray emission. The upper gray
  stripe indicates the expected contribution of LMXBs, which is a
  power law with photon index $\Gamma=1.6$ and normalization
  corresponding to the total 0.5--7~keV emissivity of LMXBs with
  luminosities $L_{\rm x}>10^{36}$\lum\ calculated using the
  analytical approximation of the LMXB luminosity function from
  \cite{gilfanov04}. The thickness of this stripe indicates the
  scatter between the normalizations of the luminosity functions of
  different galaxies \citep{gilfanov04}. The lower thick stripe shows
  a rough spectral model of the unresolved X-ray emission produced by
  weak point sources (see text), whose normalization is chosen to be
  the average over the three galaxies presented in
  Fig.~\ref{ratios}. The thickness of the curve represents the scatter
  between the emissivities of the unresolved X-ray emission of those
  galaxies.  }
\label{three}
\end{figure*}

We argued above that the X-ray emissivities of the weak X-ray
sources associated with the (old) stellar populations of the three
nearby early-type galaxies/bulges (NGC3379, M31, M32) 
are mutually compatible. However, it is natural to expect some
variations in this emissivity. First, since the main mechanism of
X-ray production in coronally active stars and cataclysmic variables
is optically thin plasma emission, metal lines are expected to provide
a considerable contribution in the energy band 0.5--2~keV (e.g. the
blend of lines at $\sim$0.8~keV). Hence the X-ray emissivity of a
stellar population should depend on its metallicity.

Second, the age of a stellar population can affect its X-ray
emissivity. For example, the X-ray luminosity of chromospherically
active stars is known to be determined by the star's rotational
velocity \cite[see e.g][]{walter81}, which in turn may depend on
stellar age. Single stars gradually slow down as they age, which leads
to a monotonic decrease of their X-ray luminosity. The evolution of
chromospherically active stars in binaries is significantly more
complicated \cite[see e.g.][]{eker06}, making it difficult to predict
the evolution of their X-ray luminosity.
Also, the orbital periods of accreting white dwarf binaries (which dominate
the unresolved X-ray emission of galaxies at energies $>$3--5 keV)
decrease with time (until they reach the so-called period minimum, see
e.g. \citealt{ps81}). This is accompanied by a decreasing mass
accretion rate \cite[e.g.][]{rjw82}. Therefore, one might 
anticipate the X-ray luminosity of CVs to decrease with age.

Despite these possible effects, the $L_{\rm x}/M_\ast$ ratios found
for the unresolved X-ray emission in M32, NGC 3379 and in the bulge of
M31 agree with each other within a factor of 2. This suggests that the
X-ray emissivity (per unit stellar mass or per unit near-infrared
luminosity) averaged over these early-type galaxies/bulges may be
regarded as a nearly universal property characterizing the old ($\ga1$ Gyr) 
stellar
population in all galaxies: $L_{\rm
  0.5-2~keV}/M_\ast=(7.0\pm2.9)\times10^{27}$ \lum\ $M^{-1}_{\odot}$
and $L_{\rm 0.5-2~keV}/L_K=(5.9\pm2.5)\times10^{27}$
\lum\ $L^{-1}_{\rm K,\odot}$. These values are the mean and the rms
scatter of the values of X-ray emissivity, measured in these three
galaxies. It is important to note however that in regions of active star 
formation the cumulative X-ray emissivity of weak stellar-type sources may be
significantly different from these values.

Since also the brighter X-ray sources belonging to the old stellar
population -- LMXBs -- are characterized by similar universality
(e.g. \citealt{gilfanov04}), we might expect the X-ray emission
of any early-type galaxy to include a resolvable (in nearby galaxies) 
component due to LMXBs and an unresolved component due to weak point
sources, with the relative contributions of these components not varying
significantly from one galaxy to another. In addition, there may be
truly diffuse emission produced by hot interstellar gas.

To illustrate this point, we compare in Fig.~\ref{three} the X-ray
(0.5--7~keV) spectra of three elliptical galaxies with very different
stellar masses and gas content. Specifically, we compare
the total {\em Chandra} spectra of M32 ($M_\ast\sim10^9
M_\odot$), NGC~3379 ($M_\ast\sim 6\times 10^{10} M_\odot$) and the
giant galaxy M87 ($M_\ast\sim4\times10^{11} M_\odot$). For M32 and
NGC~3379 we also show the spectra of their unresolved X-ray emission.  

The total emission of M32 and NGC~3379 is clearly dominated by LMXBs
and can be approximately described by a power-law spectrum with photon
index $\Gamma=1.6$, a value typical of LMXB spectra in the 0.5--7 keV
energy band \citep{irwin03}. The unresolved X-ray emission of these
galaxies, which we believe is the superposition of weak point sources,
CVs and ABs, has a spectrum that can be roughly approximated (using
the {\em Chandra} spectrum of NGC~3379) by a combination of thermal
plasma emission ($mekal$ model in XSPEC) with temperature $kT=0.5$ keV
and a Solar \citep{solar} abundance of heavy elements, and a power law
with $\Gamma=1.9$.  The ratio of the luminosities of the $mekal$ and
$powerlaw$ components in the 0.5--2 keV band was set to $2.03$.  These
components add up so that their combined emissivities in the 0.5--2
keV and 2--7 keV bands are $L_{\rm
  0.5-2~keV}/M_{\ast}=(8.2\pm2.5)\times10^{27}$ \lum\ $M^{-1}_\odot$
and $L_{\rm 2-7~keV}/M_{\ast}=(6.3\pm2.5)\times10^{27}$
\lum\ $M^{-1}_\odot$, respectively.

In the case of the giant galaxy M87 in the Virgo cluster, most of the
X-ray emission comes from the hot gas, while LMXBs and weak X-ray
sources are expected to provide $\sim 1$\% and $\sim0.1$\% of the total
emission in 0.5-2 keV energy band (see Fig.~\ref{three}).

\section{Broad energy band picture for early-type galaxies: 0.5-100
  keV}

The above discussion was restricted to the {\em Chandra} working
energy band (0.5--7~keV), with particular focus on the soft band
(0.5--2~keV) where {\em Chandra} is most efficient. We now wish to
extend this discussion to the study of early-type galaxies at 
photon energies up to $\sim$100~keV.

As was already discussed in the preceeding section, the X-ray emission
of an early-type galaxy is expected to consist of three main
components: (i) emission from LMXBs, which may be at least partly
resolved (depending on the instrument, exposure and distance to the
galaxy), (ii) unresolved diffuse-like emission produced by cataclysmic
variables and coronally active stars and (iii) diffuse thermal
emission of hot interstellar gas. There may also be additional
contributions from e.g. nonthermal emission associated with cosmic
rays or from X-ray sources representing young stellar populations (in
particular HMXBs).

Below we discuss how the relative contributions of the main
X-ray emission components of an elliptical galaxy vary across the broad
energy band 0.5--100~keV and for illustration consider the X-ray spectrum
of the giant gas rich elliptical galaxy M87 (Fig.~\ref{m87_broad}). 

\subsection{LMXBs}

Low-mass X-ray binaries typically contribute at the level of $L_{\rm
x}/M_\ast\sim8\times10^{28}$ \lum\ M$^{-1}_\odot$ in 0.5-7 keV energy band
\citep[e.g.][]{gilfanov04} and have very hard spectra at energies
$<10$ keV. At energies above 10~keV, the spectra of the brightest LMXBs
($L_{\rm x}\ga 10^{37}$~erg~s$^{-1}$), which contribute the most to
the cumulative emissivity of LMXBs at $E<10$~keV, have an exponential cutoff
\cite[e.g.][]{mitsuda84,pavlinsky94}. Therefore,  at still higher
energies ($>20$ keV) the less luminous LMXBs in the so-called low/hard
spectral state dominate. 

We show in Fig.~\ref{m87_broad} an expected broad-band spectrum of an
entire galactic population of LMXBs. It is constructed of the {\em
  Chandra} spectrum of NGC~3379 at $E<7$~keV and the cumulative
spectrum of the Galactic Center region (with the total mass of the
enclosed stellar population $M_\ast\sim 10^{10} M_\odot$) obtained by
\cite{krivonos07} with {\em RXTE}/{\em PCA} and {\em INTEGRAL}/{\em
  IBIS} at energies between 5 and 100~keV. Both parts of the spectrum
were normalized by the corresponding stellar masses and the
high-energy part was additionally multipled by a factor of 4 to
match the data points at $E<7$ keV. This additional rescaling can be
justified partly because LMXBs are somewhat less abundant in the Milky
Way than in nearby ellipticals \cite[see e.g.][]{gilfanov04} and also
because the observed number of sources which contribute the most to
the cumulative flux in the Galactic Center region explored by
\cite{krivonos07} is small. At energies 3--20~keV, it is LMXBs with
$L_{\rm x}\sim 10^{37.5}$--$10^{38}$~erg~s$^{-1}$ (3 sources), and
at energies 20--200~keV it is LXMBs with $L_{\rm x}\sim
10^{36}$--$10^{37}$~erg~s$^{-1}$ (3--4 sources).

\subsection{CVs and ABs}

In Fig.~\ref{m87_broad} we also show a composite spectrum of a
galactic population of weaker point sources, including cataclysmic
variables and coronally active binary stars. This spectrum is 
constucted from the {\em Chandra} spectrum of the unresolved emission
of NGC~3379 at energies below 7~keV and of the high-energy
(3--100~keV) spectrum of the Galactic Ridge X-ray Emission measured by
{\em RXTE}/{\em PCA} \citep{mikej06} and {\em INTEGRAL}/{\em IBIS} 
\citep{krivonos07}. This composite spectrum can be approximately
described by a $mekal$+$powerlaw$ model described in Sect.~4 with
the multiplicative addition of a high-energy exponential cutoff,
$\exp(-E/18~{\rm keV})$. 

\subsection{Hot interstellar gas}

\begin{figure}
\includegraphics[width=\columnwidth]{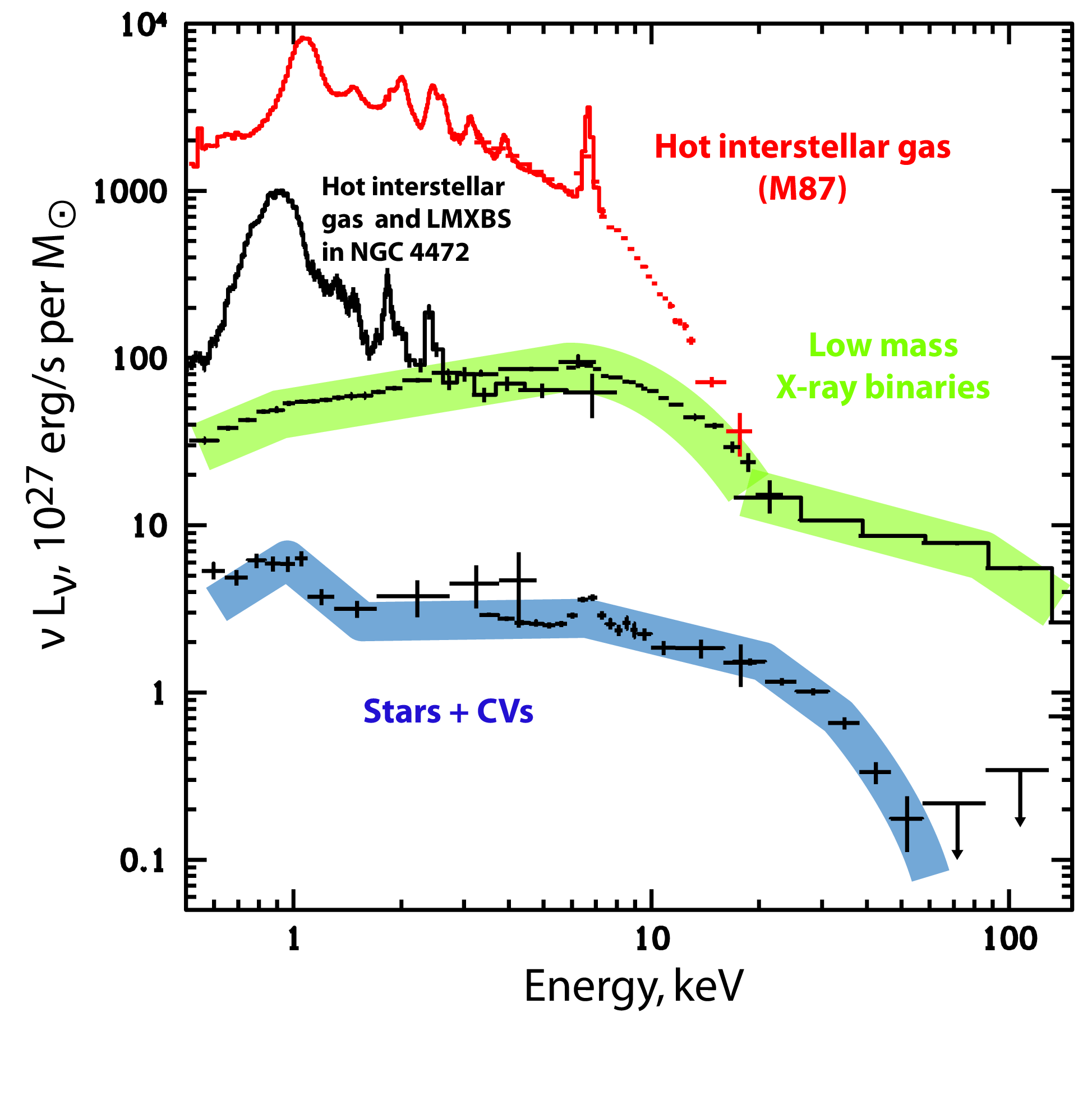}
\caption{Illustration of the composition of the X-ray spectrum of an
  elliptical galaxy from three main components (from the top to the bottom)
 -- hot interstellar gas 
(spectra of M87 and NGC~4472); cumulative spectrum of emission of low 
mass X-ray binaries (spectrum of NGC~3379 and Milky Way, see text for 
details); cumulative spectrum of active stars and CVs (spectra of
NGC~3379 and the Milky Way, see text for detals).
}
\label{m87_broad}
\end{figure}

More massive elliptical galaxies often reside in clusters or groups of
galaxies and have large reservoirs of hot gas, which produces most of
the X-ray emission of such galaxies. The best studied example is the
giant elliptical galaxy M87 in the central part of the Virgo
cluster. In Fig.~\ref{m87_broad} we show the spectrum of M87 collected
by {\em Chandra} in the 0.5--7~keV energy band within the optical
effective radius of the galaxy ($\sim$100\arcsec), to which we
attached the spectrum observed by {\em RXTE}/{\em PCA} in the
3--20~keV band within the central 1$^\circ$-radius region of M87. To
obtain a good match of the {\em Chandra} and {\em RXTE} data points,
we multiplied the latter by a factor of 0.06, which is approximately
the ratio of the X-ray luminosities of the central 100\arcsec and
1$^\circ$ regions of M87 \cite[e.g.][]{boehringer94}. Note that in the
large region the emission is dominated by the thermal gas even more
strongly than within the central 100\arcsec\ radius circle. Therefore, when
the {\em RXTE} spectrum is renormalized to match the {\em Chandra}
spectrum at energies of a few keV, the contribution of the stellar
population to the resulting spectrum shown in Fig.~\ref{m87_broad} is
smaller than one would expect for the central 100\arcsec\ region.

From the above discussion and
Fig.~\ref{m87_broad} we can draw the following conclusions:
\begin{itemize}

\item
The 0.5--100~keV emission observed from a gas poor elliptical galaxy
by a non-imaging (or poor angular resolution) instrument, such as {\em
RXTE}/{\em PCA}, will be dominated by LMXBs, although weaker X-ray sources
such as CVs and ABs will provide an important contribution (10--20\%)
in the soft X-ray band (below 2~keV) and also near 20~keV. 

\item
Future hard X-ray imaging instruments such as Nustar and Symbol-X
should be able to resolve individual LMXBs in nearby galaxies. In this
case, in gas poor elliptical galaxies there will be a residual
diffuse-like hard X-ray component due to unresolved cataclysmic
variables (coronally active stars provide a minor contribution at
energies above 10~keV) resembling emission of diffuse thermal gas with
a temperature of $\sim$20--30~keV.
 
\item
In gas rich galaxies like M87,  thermal emission of hot
intergalactic gas strongly dominates at energies below
10~keV. At higher energies, however, the contribution of point sources
can be significant or even dominant. In fact, as one can see from
Fig.~\ref{m87_broad}, LMXBs start to dominate in M87 already at
$\sim$20~keV, and if these LMXBs can be resolved by a future hard X-ray
telescope then the remaining unresolved emission at energies
20--40~keV may have comparable contributions from hot interstellar gas
and weak point sources (mostly CVs). Searches for
nonthermal hard X-ray emission from brightest cluster galaxies, like
M87, must account for these other components.

\end{itemize}

\section{Summary and the future}

In this paper we have demonstrated that the X-ray emission of
the gas poor elliptical galaxy NGC~3379 that remained unresolved after
subtraction of bright X-ray sources (predominently LMXBs) can be
explained as the cumulative emission of yet weaker descrete stellar-type
sources, namely coronally active stars and cataclysmic 
variables. The correspondence of the measured X-ray emissivity (per
unit stellar mass or per unit $K$-band luminosity) with the cumulative
emissivity of such objects in the Solar vicinity indicates that
there is no need for an additional contribution from hot interstellar
gas in this galaxy. 

The consistency (within a factor of 1.3) of the X-ray emissivities of
faint discrete sources in the galaxies M32, Milky Way, M31 and NGC~3379,
which differ in mass by more than a factor of 50, suggests that
stellar-type sources (apart from bright LMXBs) in a given early-type
galaxy of mass $M_\ast$ should produce a soft X-ray luminosity $L_{\rm
  0.5-2~keV}\sim 7\times10^{38}(M_\ast/10^{11}
M_\odot)$~\lum. Furthermore, since the majority of the contributing
sources (CVs and ABs) are expected to be relatively old ($\ga$1 Gyr),
they should provide a similar contribution also in late-type
galaxies. However, in the latter case there might be an additional
contribution from weak X-ray sources belonging to  younger stellar
populations, e.g. from rapidly rotating single coronal stars, as
indicated by observations of the Solar vicinity \citep{sazonov06}.

It is important to emphasize that the spectral shape of the emission
of faint discrete sources is very similar to the emission of
multitemperature hot interstellar gas and it is virtually 
impossible to distinguish one from the other using present-day X-ray
spectrometers. We may anticipate, however, that the next generation of
high energy resolution X-ray spectrometers will be able to
disntinguish these two types of sources by detecting the expected
redshifted lines in the emission from the accreting white dwarf in CVs
or through the expected differences in the line emission of the
relatively dense plasma in stellar coronae and in the tenuous
interstellar gas.  

We showed that the relative contributions of different constituents
to the total emission of galaxies are expected to strongly depend on
the energy band of the given study. We presented schematic broad-band
spectra of the major contributors to the X-ray emission of 
galaxies (see Fig.~\ref{m87_broad}).

The fact that the cumulative contribution of weak stellar-type X-ray sources
can provide a significant or even dominant contribution to the observed
unresolved extended X-ray emission of galaxies even as massive as NGC~3379 
($\sim6\times10^{10} M_\odot$) must be considered in different
studies such as measurements of the masses and the hot gas content of 
elliptical galaxies.

\begin{acknowledgements}
This research made use of data obtained from the High Energy Astrophysics
Science Archive Research Center Online Service, provided by the
NASA/Goddard Space Flight Center. This publication makes use of data
products from the Two Micron All Sky Survey, which is a joint project
of the University of Massachusetts and the Infrared Processing and
Analysis Center/California Institute of Technology, funded by the
National Aeronautics and Space Administration and the National
Science Foundation. This work was supported by DFG-Schwerpunktprogramme
(SPP 1177), grants CH389/3-2, RFFI~07-02-01051,
07-02-00961, NSH-5579.2008.2 and by program of Presidium of Russian Academy of Sciences ``Formation and evolution of stars and galaxies''

\end{acknowledgements}

\end{document}